\documentstyle[12pt]{article}

\newcommand{\nc}{\newcommand}
\def\frac#1#2{{\textstyle {#1 \over #2}}}

\nc{\beq}{\begin{equation}}
\nc{\eeq}{\end{equation}}
\nc{\beqa}{\begin{eqnarray}}
\nc{\eeqa}{\end{eqnarray}}
\nc{\lsim}{\begin{array}{c}\,\sim\vspace{-21pt}\\< \end{array}}
\nc{\gsim}{\begin{array}{c}\sim\vspace{-21pt}\\> \end{array}}

\newcommand{\mysection}[1]{\setcounter{equation}{0}\section{#1}}

\def\psil{\psi_L}
\def\bpsil{\bar{\psi}_L}
\def\phl{\phi^L}
\def\phr{\phi^R}

\def\&{and}

\def\DS {D\!\!\!\!/}
\def\A { A_\mu (x) }
\def\AS {A^*_\mu (x)}
\def\U { U_\mu (x) }

\def\ba{ \bar{a} }
\def\det{ det[ ~\DS_L~ ] }
\def\app#1#2#3{          {\it Acta Phys. Polon. }{\bf #1}, #2 (19#3)}

\def\com#1#2#3{          {\it Comm. Math. Phys. }{\bf #1}, #2 (19#3)}

\def\nc#1#2#3{           {\it Nuovo Cim.  }{\bf #1}, #2 (19#3)}
\def\np#1#2#3{           {\it Nucl. Phys. }{\bf #1}, #2 (19#3)}
\def\pl#1#2#3{           {\it Phys. Lett. }{\bf #1}, #2 (19#3)}
\def\pr#1#2#3{           {\it Phys. Rev. }{\bf #1}, #2 (19#3)}
\def\prep#1#2#3{         {\it Phys. Rep. }{\bf #1}, #2 (19#3)}
\def\prl#1#2#3{          {\it Phys. Rev. Lett. }{\bf #1}, #2 (19#3)}

\begin{document}
\begin{titlepage}

\begin{center}

\hfill    YCTP-P5-9   \\
\hfill  hep-th/9503064 \\
\vskip .5 in
{\large \bf Regularization of Chiral Gauge Theories}
\vskip .3 in
{
  {\bf Stephen D.H. Hsu}\footnote{hsu@hsunext.physics.yale.edu}
   \vskip 0.3 cm
   {\it Sloane Physics Laboratory,
        Yale University,
        New Haven, CT 06511}\\ }
  \vskip 0.3 cm
\end{center}

\vskip .5 in
\begin{abstract}
We propose a nonperturbative
formulation of chiral gauge theories.
The method involves a `pre-regulation' of the gauge fields, which may
be implemented on a lattice,
followed by a computation of the
chiral fermion determinant in the form of a functional integral
which is regularized in the continuum.
Our result for the chiral determinant is expressed in
terms of the vector-like Dirac operator and hence can
be realized in lattice simulations.
We investigate the local and global anomalies
within our regularization scheme.
We also compare our result for the chiral determinant to previous
exact $\zeta$-function results.
Finally, we use a
symmetry property of the chiral
determinant to show that the
partition function for a chiral gauge theory is real.

\end{abstract}
\end{titlepage}

\renewcommand{\thepage}{\arabic{page}}
\setcounter{page}{1}
\mysection{Introduction}
The non-perturbative formulation of chiral gauge theories
is a long-standing problem of quantum field theory with
both practical and theoretical implications \cite{RECENT}.
Recently there
has been a rekindling of interest, largely due
to 'tHooft \cite{TH}, in the idea of using
gauge field interpolation to couple lattice gauge fields
to continuum fermions. By keeping the fermions in the
continuum one avoids the difficulties associated with
realizing chiral fermions on the lattice \cite{NN}.
This idea was discussed previously
in the literature in \cite{FW,AG,GS} and has now
recently been further developed in \cite{AK,HS,IM}.

In this approach,
the gauge fields and fermions are treated differently,
with the gauge fields originally defined only on the links of a
spacetime lattice, as in the usual formulation of lattice gauge theory.
However, an interpolation algorithm associates the
gauge link variables $U_\mu (x)$ with a continuum gauge field
$A_\mu (x)$ in whose background the fermion part of the
functional integral is evaluated. Thus, given an interpolation
scheme the problem that remains is to give a continuum
formulation of the chiral determinant which can be
evaluated to a desired accuracy within a finite computation.

In this paper we give a simple, gauge invariant formulation of the
chiral determinant in terms of the vector-like Dirac operator which
can be realized, at least in principle, in lattice simulations.
The relevant background fields for our determinant
have been `pre-regulated' by the
lattice interpolation and for the purposes of our analysis
we will take them to be
smooth, with variation on length scales larger than a chosen
scale $\ba$, where
$\ba$ is related to the lattice spacing $a$.
We will discuss below to what extent various interpolation
schemes satisfy this property.
At the end of the procedure we allow $a, \ba \rightarrow 0$,
but only after first taking to infinity the continuum mode
cutoff $N$
used in the functional integral.
The good behavior of the background field allows us to
make well-defined manipulations of the functional integral
and in particular to separate low frequency physics from the
high frequency physics which comes from modes near
$N$.

Our partition function is
\beq
\label{Z}
Z ~=~ \sum_{  \{ U  \} } ~e^{ - S_{YM} [ U ]} ~ \det,
\eeq
where the sum is over all gauge link configurations,
the action $S_{YM}$ is the usual Yang-Mills lattice action,
and the
determinant is a functional of the {\it continuum} gauge field
$\A$ which is uniquely determined from each discrete set of
links $\{ \U \}$. The determinant
$\det$ of  the chiral Euclidean Dirac
operator in the background field $\A$ will be defined in section 3.
As we will discuss below, it is also straightforward to
define regulated fermion correlators within our scheme.

Let us recall some well-known results concerning
fermion determinants.
In Euclidean space fermion determinants for vector-like models
are real and positive semi-definite.
It is possible to convert a chiral
model with fermions in representations $r$
into a vector-like one by the addition of `mirror' fermions
which are exact copies of the originals, but
in complex conjugate gauge representations $r^*$:
\beq
det [~ \DS_V ~] ~=~ det [~ {\DS_L}^r ~] ~ det [~ {\DS_L}^{r^*} ~]
{}~=~  det [~ {\DS_L}^r ~] ~det [~ {\DS_L}^r ~]^* ~.
\eeq
Thus, we see that the
magnitude of the chiral determinant is simply equal to the
square root of the corresponding vector-like determinant.
We also see that any gauge anomaly in the chiral determinant
must be a pure phase, since a vector-like
model is anomaly free. In this paper we are primarily interested in
models in which gauge anomalies are absent, so the important
quantity is the non-anomalous
phase of $\det$, which reflects the chiral nature of the model.

Our regularized chiral determinant, described in detail in
section 3, is given  in terms of a
regularized fermion functional integral. We regulate the integral
by truncating the functional measure to a finite number of modes,
$N$.
This cutoff renders the integral finite, but at the
cost of introducing some gauge non-invariance in the
{\it magnitude} of the determinant.
Since, as mentioned above, we already know the desired magnitude
of the fermion determinant in terms of the corresponding
vector-like model, this does not lead to any ambiguity in defining the
regularized model. We simply modify the magnitude to agree
with the square root of the corresponding vector-like model.

We will argue that our determinant has the following properties:

\vspace{.5cm}

\noindent {\bf (1)} It correctly reproduces the known global and
local anomalies, as well as allowing for
anomalous phenomena such as
fermion number violation. (This is in some sense `built-in' to
our formulation as it mimics the naive continuum definition
as much as possible.)

\vspace{.3cm}

\noindent {\bf (2)}
It yields a {\it gauge invariant} result
in the limit $N \rightarrow \infty$,
at least for infinitesimal gauge transformations.

\vspace{.3cm}

\noindent {\bf (3)}  Its construction is accomplished using only
the eigenvalues and eigenfunctions of the vector-like
Dirac operator $\DS$.

\vspace{.5cm}

This paper is organized as follows. In the following section we briefly
discuss some interpolation schemes and their properties.
In section 3 we give our
definition of the chiral determinant in terms of a
regularized fermion functional integral. In section 4 we investigate
the local and global
anomalies within our regularization scheme.
We show that in models in which gauge anomalies cancel
the phase $\eta$ of our determinant is invariant under
infinitesimal gauge transformations even for finite truncation
to $N$ modes.
In section 5 we discuss how our results are related to
previously derived exact representations
of the imaginary part of $ln ( \det )$.
In section 6 we discuss a simplification of the partition function
which arises from
the behavior of the chiral determinant under reflection of
the background field.
We conclude with a summary and an
appendix on the convergence properties of the chiral determinant.

\section{Pre-Regulation of Gauge Fields}

In this section we give a brief overview of some possible
interpolation schemes, with emphasis on their smoothness
properties.  As we will see in later sections the required size of the
continuum cutoff on the fermion modes depends
on the smoothness properties of the background gauge fields.
It is desirable that the interpolated fields have their support
in momentum space concentrated at momenta less than some
scale $1 / \ba$, which is presumably controlled by and of order
the lattice spacing $a$. It is also desirable
that the interpolation scheme be gauge covariant,
so that the effect of a lattice gauge transform on the links
is consistent with effect of the interpolated gauge transform
on the continuum field. More explicitly, under a gauge transform
we should have
\beqa
\label{cov}
\{ U \} ~&\rightarrow&~ \{ U^{\Omega} \}   \nonumber \\
\A ~&\rightarrow&~  A^{\Omega_c}_{\mu} (x) ,
\eeqa
where $\Omega$ is the lattice gauge transform (valued only
on discrete lattice sites) and $\Omega_c$ its continuum interpolation.
Even given the above criteria the choice of interpolation prescription
is highly arbitrary.

The interpolation scheme based on the geometrical definition
of topological charge \cite{ML,GKSW}
obeys the gauge covariance condition, but both it and the
scheme based on minimizing the Euclidean action
proposed in \cite{TH} produce continuum fields which
are in general only piecewise continuous. Without some additional
smoothing, these configurations possess Fourier transforms
with support that falls off at large momentum $k$ only like
one over $\vert k \vert$ to some power. In general we would prefer
configurations which are completely smooth (infinitely
differentiable) and hence by the Riemann-Lebesgue lemma
have Fourier transforms that fall off exponentially
at large momenta.

A very simple scheme which leads to smooth continuum
fields was recently given by Montvay \cite{IM}.
(Unfortunately this scheme is specific to $U(1)$ gauge
theories.)
First one imposes a gauge fixing such as Landau gauge to
define a set of gauge fields $A_{x \mu}$ on the discrete
spacetime lattice. (This can always be done in such a way that
the $A_{x \mu}$ are bounded by $2 \pi / a$.)
The Fourier coefficients $\tilde{A}_{k \mu}$
of the lattice gauge field $A_{x \mu}$
are valued only on discrete values of $k_\mu$ in the
Brillouin zone:  $k_{\mu} = 2 \pi n_{\mu} / L_{\mu}~$,
$~0 \leq n_\mu \leq L_\mu /a$.
($L_\mu$ is the extent of the lattice in the $\mu$ direction.)
One can use these Fourier coefficents  to define a continuum
gauge field through
\beq
\label{smooth}
\A ~=~ \frac{1}{ ( \prod_\nu L_\nu ) } ~
\sum_{k_\mu} e^{ik \cdot x} ~ \tilde{A}_{k \mu}.
\eeq
The resulting continuum gauge field is infinitely differentiable, and
has zero support in momentum space for $\vert k \vert > 2 \pi / a$.
A slightly modified version of this scheme can be given \cite{IM}
which also satisfies the covariance
condition (\ref{cov}). An interpolation like (\ref{smooth})
relates the lattice and continuum gauge transforms and
guarantees that the latter also have support only in the
Brillouin zone.

For our purposes in what follows, we will assume that the
gauge field backgrounds in which our functional determinant
is to be evaluated are bounded and infinitely differentiable,
and hence that
their Fourier transform falls off exponentially rapidly at
momenta large compared to a scale $1 / \ba$ which is
controlled by the lattice spacing $a$.
We will also restrict ourselves to smooth gauge transforms
$\Omega (x)$, such as would result from an interpolation
of the type above from a lattice gauge transform.
This implies that in the
band-diagonal matrices which will appear in the next section,
the entries which are outside the bands can be made zero
by choice of the width of the band. In the specific case of an
interpolation like that of \cite{IM}, the width of the bands will
be $\sim 1/a$ with entries outside the bands exactly zero.

We should note that while this assumption of smoothness of
the background field allows us to discern various nice
features of the determinant, such as the band-diagonal
properties of certain matrices, it is not absolutely necessary to
demonstrate the three main properties listed in the introduction.
For that purpose the interpolations of \cite{ML,GKSW,TH}
with their piecewise continuity should suffice.

\renewcommand{\thepage}{\arabic{page}}
\mysection{Chiral Determinant}

Our task is now to define the functional determinant in
the smooth background $\A$. We will follow a straightforward
procedure similar to that first used
by Fujikawa \cite{FU} in his
functional integral approach to the anomaly. We work in
Euclidean space-time which results from the Wick rotation
$x^0 \rightarrow -i x^4$ and $ A_0 \rightarrow i A_4$.
The Dirac operator
$\DS ~\equiv~ \gamma^\mu D_\mu ~=~
\gamma^{\mu} ( \partial_\mu + A_\mu )$ becomes
a Hermitian operator
\beq
\label{Dirac}
\DS ~=~ \gamma^4 D_4 ~+~ \gamma^i D_i  .
\eeq
We use the convention that $\gamma^0$ is Hermitian and
$\gamma^k$ anti-Hermitian. The Hermitian $\gamma^5$ is
defined as
\beq
\gamma_5 ~\equiv~ i \gamma^0 \gamma^1 \gamma^2 \gamma^3
           ~=~ - \gamma^1 \gamma^2 \gamma^3 \gamma^4.
\eeq
The Euclidean metric is $g_{\mu \nu} ~=~ (-1, -1, -1, -1)$.

The determinant is defined, formally, through the fermion
functional integral
\beq
\label{FI}
\det~\equiv~ \int D \psil~ D \bpsil~
  e^{ i \int d^4x \bpsil \DS \psil }.
\eeq
The chiral fields are given in terms
of the complete set of orthonormal
eigenfunctions of the Hermitian Dirac operator:
\beq
\label{eigen}
\DS ~ \phi_n ~=~ \lambda_n~ \phi_n.
\eeq
As usual, we imagine that our system has
been placed in a box of size
$L$ with appropriate boundary conditions
in order to discretize
the eigenvalues.
We define a complete set of chiral modes by
\beqa
\label{modes}
\phi_n^L ~&=&~ \sqrt{2} P_L~ \phi_n ~~~~~~ \lambda_n > 0
\nobreak\\ \nonumber
    &=&     P_L   \phi_n ~~~~~~~~~~~~ \lambda_n = 0,
\eeqa
where $P_L ~=~( \frac{ 1 - \gamma_5}{2} )$.
The modes $\phr_n$ are defined equivalently but with left-handed
projectors replaced with
right-handed ones.
The basis $ \{ \phl_n, \phr_n  \}$ is complete and orthonormal,
as we can see
since $\DS~ \gamma_5 \phi_n ~=~ - \lambda_n~\gamma_5 \phi_n$.

A subtle but important point should be emphasized here:
the choice of positive
eigenvalue modes which are used to define
$\{ \phl_n, \phr_n  \}$ must be determined
for some {\it fiducial} value of $\A$, which
we will take to be $\A = 0$. As the gauge field is varied
from zero to some arbitrary configuration $\A$,
the evolution of modes which initially had positive
eigenvalues $\lambda_n \geq 0$ must be followed in
order to define the new set of modes in the
$\A$ background. In the process, some of the initially
positive eigenvalue solutions may end with
negative eigenvalues. It is this phenomena that is
responsible for possible sign changes of
the chiral determinant\footnote{In practice, this
tracking of eigenmodes would be a rather cumbersome
procedure, requiring an interpolation of the background
field of interest to $\A = 0$ and the computation of the
low-lying eigenmodes over this interpolation.}.

If we had not followed the procedure of `tracking'
the modes from the fiducial background to the
background of interest, but rather applied the
definition (\ref{modes}) in a naive way at each
value of $\A$, discontinuities in functional derivatives such as
$\frac{\delta} {\delta \A} \det$ could arise.
This sign ambiguity in the definition of
the determinant is the cause of the well-known
global anomaly, as first described by
Witten \cite{ED} \footnote{Fujikawa \cite{FU}
is primarily interested in the effect of infinitesimal
transformations on the functional measure.
His formulation, which ignores the tracking of
eigenmodes, is sensitive only to local anomalies.}.
We will return to this point
in the next section when we show that our
regularization of the determinant reproduces
the correct results for the global anomaly.

Having chosen an orthonormal basis for the background,
we can then expand
\beqa
\label{modexp}
\psil (x) ~\equiv~ \sum ~a_n \phl_n (x)    \nonumber \nobreak \\
\bpsil (x) ~\equiv~ \sum ~\bar{b}_n
{\phr_n}  (x)^{\dagger} .
\eeqa
With this expansion, the functional integral
takes on a particularly simple
form
\beqa
\label{FI1}
\det ~&\equiv&~ \int~ \prod_n \prod_m d\bar{b}_n da_m ~
e^{ ~i ~\sum_n \lambda_n \bar{b}_n a_n }
{}~det [~C^L~] \cdot det [~C^R~]\nonumber \\
&=& ~det [~C^L~] \cdot det [~C^R~] ~\prod_n ~i~\lambda_n  ~ .
\eeqa
Fermion correlators can also be explicity evaluated, for
example, the propagator
\beqa
\label{prop}
\langle  \bpsil (x) \psil (0) \rangle &  ~=~ &
\int D \psil~ D \bpsil~ ( \bpsil (x) \psil (0) )~
  e^{ - \int d^4x \bpsil \DS \psil }   \nonumber  \\
 & ~=~ &
 \left( \sum_m ~\frac{1}{\lambda_m}~
\phi_m^R (x)^\dagger \phi^L_m (0)
\right) ~
{}~det [~C^L~] \cdot det [~C^R~]~
\prod_n ~i~\lambda_n  .
\eeqa
The factor $\prod_n \lambda_n$ is simply the square root of
the corresponding vector-like
functional integral
(i.e. with no chiral projector in the action of (\ref{FI}) ), up to
the sign ambiguity we discussed previously.
The factors of $i$ in the infinite product amount to an overall
phase factor which is $\A$ independent and are
irrelevant to our analysis. We will drop them in what follows.

The additional Jacobian factors $det [~C^L~]~,~det [~C^R~]$
arise from the change in basis
we have had to make from an initial {\it fiducial} basis which
we define as a product over {\it free} modes.
In other words, we choose an initial fiducial measure
for the functional integral
\beq
\label{FM}
\prod_n \prod_m~ d\bar{b}_n da_m,
\eeq
where the coefficients $\bar{b}_n, a_n$ are those that
arise in the expansion
of (\ref{modexp}) in terms of {\it free} solutions.
In order to perform the integral using the
orthonormality properties of solutions in
the $\A$ background, we have to change basis
and hence the extra Jacobian factors must appear.

The matrices $C^L, C^R$ are defined as follows. Let the lack of an additional
superscript denote free eigenmodes
and the superscript $A$ denote eigenmodes in the $\A$ background. Then
\beqa
\label{CL}
C^L_{mn} ~&\equiv&~  \langle \phi^{L,A}_m \vert \phl_n  \rangle \nonumber \\
{}~&=&~  \int d^4x~ \phi^{L,A}_m (x) ^{\dagger} \phl_n (x) \nonumber \\
            ~&=&~  \int d^4x~ \phi_m^A (x) ^{\dagger}~ P_L ~\phi_n.
\eeqa
Similarly,
\beqa
\label{CR}
C^R_{mn}  ~&\equiv&~  \langle \phi^{R}_n \vert
\phi^{R,A}_m \rangle    \nonumber \\
{}~&=&~	\int d^4x~ \phi^{R}_n (x) ^{\dagger} \phi^{R,A}_m (x) \nonumber \\
    ~&=&~  \int d^4x~ \phi_n (x) ^{\dagger}~ P_R ~\phi^A_m (x).
\eeqa

The $C$ matrices are complex but unitary, so the
factor $det [~C^L~] \cdot det [~C^R~]$ is formally a pure phase.
This phase, when combined with the potential sign changes in
$\prod_n ~\lambda_n$ (equivalent to phases of $i \pi$),
constitute the chiral phase information mentioned in the introduction.
In the absence of gauge
anomalies,
$det[ ~\DS_L~ ]$ should be
gauge invariant, which in turn requires that
the result (\ref{FI1}) is gauge invariant.
At the formal level, a gauge transformation
acts as a unitary transformation on the
$C^{L,R}$ matrices, and hence
should leave
$det [~C^L~] \cdot det [~C^R~]$ invariant.
We will examine the gauge invariance properties
of our regulated version of this object in section 4.

In what follows we will regularize
all of our previous expressions by eliminating
all but a finite number of eigenmodes from
our functional integral.
This will truncate all
infinite sums and products to finite ones, and
also infinite matrices to finite matrices.
All expressions and manipulations will then
be well-defined and finite, and could in princple
be implemented on a gedanken-computer.
(See \cite{AK,AB} for work on
similar regularization schemes.)
Of course, this regularization is precisely a
`hard cutoff' in the space of eigenfunctions, and
violates gauge invariance. One of our main
results will be that for pre-regulated backgrounds
$\A$, the violations of gauge invariance are
limited and readily compensated.
In particular, gauge non-invariance due to the
finite truncation will be seen to only affect
the magnitude of the chiral determinant,
allowing the chiral phase information
to be extracted in a well-defined manner.
Our truncated expression for $\det$ is then
\beq
\label{detreg}
det [~C^L~] \cdot det [~C^R~] ~ \prod_n^N~ \lambda_n  ~ ,
\eeq
and for the propagator
\beq
\label{regprop}
\langle  \bpsil (x) \psil (0) \rangle   ~=~
 \left( \sum_m^N ~\frac{1}{\lambda_m}~
\phi_m^R (x)^\dagger \phi^L_m (0) \
\right) ~
{}~ det [~C^L~] \cdot det [~C^R~] ~
\prod_n^N  \lambda_n  .
\eeq
where the $C$ matrices are now finite dimensional.
The number of modes that are kept is
$N ~\sim~  (L \Lambda)^4 $, where
$\Lambda$ is roughly the UV scale associated
with our mode truncation.
(This is up to additional factors due to
degeneracies and zero modes.)
Here $L$ is the size of our box.
As we will specify, the scale at which
our truncation is made is determined
by the smoothness scale $\ba$, the lattice spacing $a$ and
the box size $L$.

The pre-regulation (smoothness) of the gauge field
implies that the $C$ matrices have a very simple
form. We can see this by first
making a useful observation
about the explicit forms of the modes which
appear in (\ref{CL}) and (\ref{CR}).
Because the free basis is
essentially a plane wave basis, it is useful to examine our
eigenvalue equation (\ref{eigen}) in momentum space.
The eigenvalue equation for
$\phi_n (q)$, has the form
\beq
\label{qeigen}
( i q\!\!\!/ - \lambda_n) \phi_n (q) ~+~
\int d^4k~ A\!\!\!/  (q-k) \phi_n(k) ~=~ 0.
\eeq
Here $A_\mu (k)$ represents the Fourier
transform of the background field $\A$.
Because of the smoothness property of $\A$, the Riemann-Lebesgue
lemma tells us that $A(k) \rightarrow 0$ as
$\vert k \vert \rightarrow \infty$. In fact, $A(k)$
goes to zero exponentially rapidly for
$\vert k \vert  \ba >> 1$.

We will now show that the solution to
(\ref{qeigen}), $\phi_n (q)$, has its support only in regions
of momentum space centered around values of $q$ which
satisfy $q^2 ~=~ \lambda_n^2$.
The size of those regions of support is of course
determined by the properties of $\A$.
To simplify (\ref{qeigen}), let us choose the Dirac basis for
our gamma matrices so that $\gamma_4$ is diagonal, and
the rest are off-diagonal. Now let us choose a frame in which
the momentum $q_i ~=~ ( 0, 0, 0, q_4)$. We will show that if
$q_4$ is sufficiently different from $ \pm \lambda_n$, there
is no solution to (\ref{qeigen}).

Let the the four-spinor $\phi_n (q)$ have the two-spinor components
$u_n (q), v_n (q)$.
In the frame we have chosen, the eigenvalue equation becomes
\beqa
\label{qe1}
(q_4 ~+~ \lambda_n) u_n (q) ~&=&~  \int d^4k~ \left(
i A_4 (q-k) u_n (k) ~+~ A_i (q-k)
\sigma_i v_n (k) \right) \\
(q_4 ~-~ \lambda_n) v_n (q) ~&=&~  \int d^4k~ \left(
i A_4 (q-k) v_n (k) ~+~ A_i (q-k) \sigma_i u_n (k) \right).
\eeqa

Multiply the top equation in (\ref{qe1}) by $u_n^{\dagger} (q)$
and the bottom by $v_n^{\dagger} (q)$ and integrate both over a
ball $B$ of  size $\ba^{-1}$ centered about our chosen
$q^*_i ~=~ (0,0,0,q^*_4)$  (so $q ~=~ q^* ~+~ q'$, and we
integrate $\int_B d^4q'~$
with $\vert q' \vert ~<~ \ba^{-1}$).

We can then rewrite the equations as
\beqa
\label{qe2}
(q^*_4 ~-~ \lambda_n) ~&=&~ \cdots \nonumber \\
(q^*_4 ~+~ \lambda_n) ~&=&~ \cdots ~,
\eeqa
where the terms on the right hand side denoted by $~\cdots~$
are bounded, as we will see below.
This implies that there is no solution when
$\vert q^*_4 \pm \lambda_n \vert$ are both taken
sufficiently large.

The terms on the right hand side are of three types.
In matrix notation, the first
type is of the form
(we suppress numerical factors like $\pi$)
\beq
\label{Amat}
{ \langle  u \vert A  \vert u \rangle
\over \langle  u \vert  u \rangle }
{}~<~ (L / \ba)^2 a^{-1}.
\eeq
To compute the bound in (\ref{Amat})
we have used Parseval's theorem,
\beq
\int d^4 k~ \vert A(k) \vert^2 ~=~
\int d^4x~ \vert A(x) \vert^2 ~<~ L^4 a^{-2}
\eeq
and the result that $A (k)$ has support only for momenta
$\vert k \vert \ba \lsim 1$.
The second type of term is
\beq
\label{Amat1}
{ \langle  u \vert A \vert v \rangle
\over \langle  u \vert  u \rangle }
{}~<~ (L / \ba)^2 a^{-1},
\eeq
where the bound applies if $u_n$ and $v_n$ both
satisfy (\ref{qe1}) and for $q^*_4$ sufficiently different
from both $\pm \lambda_n$. (If $u_n$ and $v_n$ were left
arbitrary, one could choose
$\vert u_n \vert \rightarrow 0$ while $v_n$ remains fixed,
which would make the left hand side of
(\ref{Amat1}) arbitrarily large.
However, it is easy to see that such $u_n$ and $v_n$ cannot
satisfy (\ref{qe1}) when the
lhs of those equations are sufficiently large.)

The last type of term is of the form
\beq
\int_B d^4 q' ~ q' ~u_n^{\dagger} (q) v_n(q) \over
\int_B d^4 q' ~ u_n^{\dagger} (q) u_n (q),
\eeq
and is bounded by $\ba^{-1}$ if again both
$u_n$ and $v_n$ satisfy (\ref{qe1}) and
$q^*_4$ is sufficiently different
from both $\pm \lambda_n$.

Thus we conclude that eigenfunctions which satisfy
(\ref{qeigen}) have support
only in regions of momentum space which
lie within a distance of order
\beq
\label{max}
{\bf max} [~ (L / \ba)^2 a^{-1}, \ba^{-1} ~]
\eeq
of values $q$ which satisfy $q^2 ~=~ \lambda_n^2$.

Intuitively, we can understand this result as follows:
the smooth background $A$ has support in a compact region of
momentum space, and is bounded in magnitude as well
$( \vert A \vert < a^{-1})$.
Its effect on eigenmodes is to `mix-up' the original
plane wave modes, but only those that are
within a certain band of each other,
whose size is determined by $A$.
The overlap between
modes $\phi_m, \phi_n$ is zero for sufficiently large
$\vert m - n \vert$. We define the
number $N_A$ such that the overlap between any modes with
$\vert m - n \vert ~>~ N_A$ is zero.

The $C$ matrices therefore have the band-diagonal
form displayed below.
We will always assume that the size of the matrix
$N$ is much greater than the width of the band $N_A$.
This roughly corresponds to a choice of cutoff $\Lambda$ for
the regularization of our
determinant. We note that from (\ref{max}) it is
clear that $\Lambda$ must always be
much larger than the corresponding lattice cutoff $a^{-1}$.
\beq
\label{CMatrix}
C^L ~,~ C^R  ~\sim~
\pmatrix{ \bullet&\bullet&\bullet&0&0 &0 &0 &0 &0 \cr
                \bullet&\bullet&\bullet&\bullet&0 &0 &0 &0 &0 \cr
                \bullet&\bullet&\bullet&\bullet&\bullet&0 &0 &0 &0 \cr
		0 &\bullet&\bullet&\bullet&\bullet&\bullet&0 &0 &0 \cr
	        0  & 0&\bullet&\bullet&\bullet&\bullet&\bullet&0 &0  \cr
		0 & 0& 0&\bullet&\bullet&\bullet&\bullet&\bullet&0 \cr
		0 & 0& 0& 0& \bullet&\bullet&\bullet&\bullet&\bullet\cr
		0 & 0& 0& 0& 0&\bullet&\bullet&\bullet&\bullet \cr
		0 & 0& 0& 0& 0& 0&\bullet&\bullet&\bullet \cr}
\eeq
For simplicity, we have adopted an index  in the above matrix
representing `momentum' k rather than the
index $n$. The latter runs from $0$ to $\infty$ whereas the momentum
can be positive or negative.
We are pretending that there is only one component of momentum --
in reality the $C$ matrices are actually multidimensional  with
additional labels representing the individual momenta
$k_i$, as well as internal group indices. The multidimensional
matrices have their band structure centered about vectors
$k_i$ in momentum
space satisfying $\sum_{i=1}^4 k_i^2 ~\sim~ \lambda_n^2$ for some
$\lambda_n$.

Now we can see the problem that arises with
truncation to a finite number of modes:
the finite dimensional matrices
$C^L, C^R$ are no longer unitary.
In fact, the products $C^L C^{L \dagger}$ and $C^R C^{R \dagger}$
are no longer the
identity but have the following form:
\beq
\label{CCMatrix}
C^L C^{L \dagger}~,~C^R C^{R \dagger} ~\sim~
\pmatrix{  \bullet & \bullet &  &  &  &  &  &  & \cr
		\bullet &  \bullet&  &  &  &  &  &  & \cr
		&  & 1 &  &  &  &  &  &		\cr
	       &  &  & 1 &  &  &  &  &   \cr
		&  &  &  & \ddots &  &  &  &   \cr
		&  &  &  &  &  & 1 &  &   \cr
		&  &  &  &  &  &  & \bullet &  \bullet \cr
		&  &  &  &  &  &  &  \bullet &  \bullet   \cr  },
\eeq
Let us denote the non-diagonal submatrices in
the upper left and lower right generically
as $X^{L,R}$. They are of dimension
$N_A$ and are the result of a loss of unitarity
that comes from our truncation. Since an infinite
number of eigenfunctions
are necessary to span the space of solutions,
unitarity requires an infinite
sum:
\beq
\label{unitarity}
\delta_{mp}
{}~=~ \sum_n^{\infty}~ C^L_{mn}  C^{L \dagger}_{np}
{}~=~ \sum_n^{\infty} ~\langle \phi^{L,A}_m \vert \phl_n  \rangle~
                 \langle \phi^{L}_n \vert \phi^{L,A}_p  \rangle.
\eeq
{}From previous analysis we know that overlaps between
modes $m$ and $n$ for which
$\vert m - n \vert > N_A$ are negligible. Therefore, for
$m, p < (N  - N_A)$
the unitarity relation (\ref{unitarity}) is unaffected.
On the other hand, for
$ (N - N_A) ~<~   m, p  ~<~ N$,
intermediate modes in the sum with nontrivial
overlap are removed in the
truncation, and hence the $X^{L,R}$ matrices
in the figure above are no longer necessarily close to the identity.
Note that the dimensionality of this matrix is independent of
the parameter $N$ and the
mode truncation in general as long as $N >> N_A$.

{}From the form of (\ref{CCMatrix}), we see that
the
product
$det [~C^L~] \cdot det [~C^{L \dagger}~]$
is now equal to:
\beq
\label{CCX}
det [~C^L~] \cdot det [~C^{L \dagger}~] ~=~ det [~X^L~],
\eeq
and similarly for the $C^R$ matrices.
The determinants of $X^{L,R}$ are real, as
$X^{L,R}$ is Hermitian:
\beqa
\label{XA}
( X^L )_{mp} ~&=&~ \sum_{n = N - N_A}^{N} ~
\langle \phi^{L,A}_m \vert \phl_n  \rangle~
                 \langle \phi^{L}_n \vert \phi^{L,A}_p  \rangle \nonumber \\
 &=& \sum_{n = N - N_A}^{N} ~ \langle \phi^{L}_n \vert \phi^{L,A}_p
\rangle ~\langle \phi^{L,A}_m \vert \phl_n  \rangle~ \nonumber \\
  &=& (X^*_L)_{pm} ~=~ (X^L)^{\dagger}_{mp} ,
\eeqa
with similar result for $X^R$.

As noted previously, the formal unitarity of the
infinite dimensional $C$ matrices
implies that $det [~C^L~] \cdot det [~C^R~]$ is a pure
phase. For finite dimensional $C$ matrices the
constraint that remains
from unitarity is that
\beq
\vert ~ det [~C^L~] \cdot det [~C^R~]~ \vert^2
{}~=~  det [~X^L~] \cdot det [~X^R~] ~\equiv~ det [~X~]^2,
\eeq
which allows for an arbitrary phase in $\det$, but does not
restrict the magnitude to be unity.
Here we define $X$ as a diagonal
matrix
$~{\rm diag} \{ \lambda^X_{1}, \cdots ,\lambda^X_{N_A} \}~$,
where the $\lambda^X_n$ are given by the square root
of the product of the corresponding n-th
eigenvalues of $X^L, X^R$.
$X$ is Hermitian
and $det[ ~X~]$ is real, so
we can still extract
the phase unambiguously from
the finite  dimensional $C$ matrices, where
\beq
\label{defeta}
det [~C^L~] \cdot det [~C^R~] ~\equiv~ e^{i \eta}~ det [~X~]
\eeq
or
\beq
\label{defeta1}
\eta ~\equiv~ {\rm Im} \left( ln \left(
det [~C^L~] \cdot det [~C^R~]  \right) \right) ~~.
\eeq

In the next section we will examine the behavior of
$det [~C^L~] \cdot det [~C^R~]$
under gauge transformations.
Any anomaly, being a pure phase,
resides in the factor $e^{i \eta}$.
In the absence of gauge anomalies, $\eta$ is
gauge invariant, which
is equivalent to the requirement that under
a gauge transformation, the change in
$det [~C^L~] \cdot det [~C^R~]$
is purely real. We will see explicitly that this
is the case in section 4.

Our final result for the regularized
chiral determinant is
\beq
\label{detfin}
\det_{reg} ~=~  e^{i \eta} ~\prod_n^N ~\lambda_n~,
\eeq
where the phase $\eta$ is gauge invariant in the absence of
gauge anomalies and
must be extracted from the finite dimensional
$C$ matrices via (\ref{defeta}) or (\ref{defeta1}).
In the appendix we discuss the convergence
properties of $\eta$. It appears that $\eta$ is determined
by the $N_A \times N_A$ submatrices of $C^{L,R}$
(i.e. it depends on the properties of eigenmodes with $n \lsim N_A$), and
therefore converges to a well-defined value as $N \rightarrow \infty$.
An alternate method of
extracting the complex phase in
(\ref{detfin}) already exists, as we will discuss in section 5.

\renewcommand{\thepage}{\arabic{page}}
\mysection{Local and Global Anomalies}

In this section we examine the local and global
anomalies within our regularization scheme.
We will verify that our regularization
scheme leads to the usual anomalous Ward-Takahashi
(WT) identities
in the limit that $N \rightarrow \infty$
and that
in the absence of anomalies the phase factor $\eta$ is
gauge invariant. Finally, we shall examine how
Witten's global anomaly is manifested in our scheme.

First let us review the behavior of the functional
measure under rotations of the fermion fields
\cite{FU}:
\beqa
\label{chirot}
\psil &\rightarrow& e^{-i \alpha (x)} \psil    \nonumber \\
\bpsil  &\rightarrow& \bpsil e^{ i \alpha (x) }.
\eeqa
These rotations can also correspond to non-Abelian
gauge transformations if we
allow $\alpha (x)$ to be an element of the Lie group:
$\alpha (x) = \alpha^a (x) T^a$.
Any subtle effects from such a transformation are to be found
in the functional measure, or equivalently
in the Jacobian determinants $det [~C^L~] \cdot det [~C^R~]$.
As usual, for infinitesimal rotations we can write
\beqa
\label{inf}
det [~C^L ~] ~&=&~ det[ ~ \delta_{mn}  +
 i \int d^4x~ \alpha (x) \phl_m (x)^\dagger \phl_n (x) ] \nonumber \\
 &=&~  exp[~ i  \sum_{n} \int~ d^4x~ \alpha (x)
\phl_n (x)^\dagger \phl_n (x) ]
\eeqa
and similarly,
\beq
det [~C^R~] ~=~ exp[ - i  \sum_{n}
\int d^4x~ \alpha (x) \phr_n (x)^\dagger \phr_n (x) ] .
\eeq
Combining these equations gives
\beqa
\label{anomaly}
det [~C^L~]~det [~C^R~] ~&=&~ exp[~ i \sum_{n} \int d^4x~ \alpha (x)
( \phl_n (x)^\dagger \phl_n (x)  ~-~
 \phr_n (x)^\dagger \phr_n (x) ) ~ ]   \nonumber \\
{}~&=&~ exp [~i \sum_{n}
\int d^4x~ \alpha (x) \phi_n (x)^\dagger  \gamma_5 \phi_n (x) ].
\eeqa
In
Fujikawa's scheme \cite{FU}, the infinite sums
are regulated by the insertion of a convergence factor of
\beq
\label{Freg}
  f (\DS^2 / \Lambda^2)   ~=~  f( \lambda_n^2 / \Lambda^2),
\eeq
where $f(0) = 1$ and
$f(\infty) = f'(\infty) = f''(\infty) ~\cdots ~=~ 0$.
The result is unchanged
as long as the function $f$ is smooth and obeys the
above boundary conditions.
The usual choice for $f (\DS^2 / \Lambda^2)$ is
$f (\DS^2 / \Lambda^2) = exp( - \DS^2 / \Lambda^2 )$.

Using Fujikawa's result in the limit $N \rightarrow \infty$
(which corresponds to taking $\Lambda \rightarrow \infty$) yields
the well-known result
\beq
\label{it}
\lim_{\Lambda \rightarrow \infty}~
\sum^{\infty}_{n}
\phi_n (x)^\dagger  \gamma_5 \phi_n (x) ~
f( \lambda_n^2 / \Lambda^2) ~=~
 - \frac{ 1}{16 \pi^2}~ F \tilde{F} (x).
\eeq
If the original transformation (\ref{chirot})
had been a non-Abelian
gauge transformation, (\ref{it}) would have additional
color structure and be proportional to
$Tr [   \{ T^a, T^b \}  T^c  ]$.

One could instead have taken the function
$f (\DS^2 / \Lambda^2)$ to approach a step-function, so
as to reproduce a hard mode cutofff \cite{AB,AK}. Since
the result is independent of the detailed form
of $f$, we obtain the usual anomalous WT identity for the
regulated current
\beq
\label{WT}
\partial_{\mu} J_{\mu} ~=~   \frac{ 1}{16 \pi^2}~ F \tilde{F} (x).
\eeq
Note that the fermionic currents $J_\mu (x)$
are themselves divergent objects  (they involve a product of
operators evaluated at
the same spacetime point $x$)
which require regularization
and some choice of subtraction in their definition.
In our scheme any correlator
is regulated automatically and non-locally by the mode truncation.
For example, see (\ref{regprop}). As $x \rightarrow 0$ the usual
short-distance divergence is cut-off by the truncation of
the series at $N$.

{}From (\ref{anomaly}) we see that the absence of anomalies
implies that the phase factor
$\eta$ defined by the $N \rightarrow \infty$ limit of (\ref{defeta1})
is gauge invariant, at least under infinitesimal rotations.
The corrections that would appear on the RHS of
(\ref{it}) due to a finite truncation are of the form,
e.g., $Tr F^k / M^{2k-4}$ ($k > 2$) and vanish as $N$ is
taken to infinity. What is more, the corrections correspond
to local operators and can be compensated by proper
choice of local counterterms if desired.

Let us now consider the global anomaly \cite{ED}, which
arises in $SU(2)$ gauge theories
due to the fact that the fourth homotopy group of
$SU(2)$ is nontrivial,
\beq
\label{homo}
\pi^4 ( SU(2) ) ~=~ Z_2.
\eeq
This means that in four dimensional Euclidean space there is a
gauge transformation $\Omega (x)$ such that $\Omega (x) \rightarrow 1$ as
$\vert x \vert \rightarrow \infty$, and $\Omega (x)$
covers the gauge group in such a way that it cannot be
continuously deformed to the
identity. The so-called mod two Atiyah-Singer index theorem \cite{APS}
then implies that as a gauge background
$\A$ is changed continuously to its gauge transform $A^\Omega_{\mu} (x) $
(e.g. via $A^t_\mu (x) ~=~ A^\Omega_{\mu} (x)
{}~+~ (1-t)~(\A ~-~ A^\Omega_{\mu} (x) )$
as $t$ goes from zero to one),
an {\it odd} number of positive-negative pairs of
eigenvalues of $\DS$ will switch places.
(The set of eigenvalues must match exactly at $t=0$ and $t=1$ since
$\A$ and $A^\Omega_{\mu} (x) $ are related by a gauge transform.)
This leads to an overall change in sign in the product
\beq
\prod_n^{\infty} ~ \lambda_n
\eeq
as long as we choose our
eigenvalues in the way we have described in the previous section,
tracking modes continuously
beginning from some fiducial background $\A$.

Now, given that we are only interested in backgrounds
which arise from our pre-regulation
procedure, it is easy to see that any modes which switch places are
within roughly $N_A$ of $n=0$. Therefore, as long as $N >> N_A$,
the truncated product
\beq
\prod_n^N ~\lambda_n
\eeq
which appears in our regulated determinant undergoes the same sign change
as the infinite product above when the background field is transformed by
$\Omega (x)$.
(Of course, this $\Omega (x)$ must relate two backgrounds
$A$ and $A'$ which satisfy
our smoothness conditions, so $\Omega (x)$ itself must be smooth.)
The modes near the cutoff, $N -N_A ~<~  n  ~<~ N$, are not
necessary to reproduce this result.

\mysection{Exact Representations}

In this section we compare our form of the determinant
to exact representations
previously obtained using $\zeta$-function methods
 \cite{AG,eta}.
We will not give the derivation of the results of
\cite{AG,eta} here, since the
details are somewhat technical, but will merely state them.
Suppose we wish to compute the imaginary part of
the chiral effective action
$ln ( \det )$ in the background $\A$ relative to
$ln ( \det )$ in the fiducial
background $\A = 0$ (we take the latter to have no phase as a
choice of convention). First form the five dimensional background
gauge field $A_t$ which interpolates adiabatically between
$\A = 0$ and our chosen $\A$ as the parameter $t$ varies from
$- \infty$ to $+ \infty$. Next,
consider the 5-dimensional Dirac operator defined by
\beq
\label{D5}
\DS_5 ~=~ ( ~ i \gamma^5 \partial_5 ~+~ \DS_4 ~),
\eeq
where $\DS_4$ is the vector-like Dirac operator in four dimensions.
The result in an anomaly free model is the
following:
\beq
\label{imdet}
{\rm Im}~ ln( \det ) ~=~ \pi ~(  ~\eta (0) ~+~ {\rm dim~ ker}~ \DS_5 ~)  ,
\eeq
where $\eta (0)$ is the famous `$\eta$-invariant'
of the Atiyah-Singer
Index theorem. It is given by the analytic
continuation to $s = 0$ of
\beq
\label{eta}
\eta (s) ~=~ \sum_{\lambda \neq 0}
\frac{ {\rm sign} ( \lambda )}
{ \vert \lambda \vert ^{-s} },
\eeq
where $\lambda$ denotes eigenvalues of $\DS_5$.
One can think of
$\eta (0)$ as the regularized
spectral asymmetry of $\DS_5$:
\beq
\label{eta1}
\eta (0) ~\sim~ \sum_{\lambda > 0} 1 ~-~ \sum_{\lambda < 0} 1.
\eeq

The term
$~ {\rm dim~ ker}~ \DS_5~$ simply counts the number of zero modes
of the operator $\DS_5$. We see that the determinant changes
sign whenever such a zero mode appears.
An identical change in sign can be seen to result in our treatment
if we recall Witten's result \cite{ED}, using a construction
like the five dimensional one above, that  a zero mode of $\DS_5$
implies that in the
interpolated background $A_t$ the  flow of eigenvalues of $\DS_4$ is
such that an {\it odd} number of eigenvalue pairs change sign.
Thus the sign change in due to $~ {\rm dim~ ker}~ \DS_5~$
in the exact representation matches that due to eigenvalue flow
in our form of the determinant.

This leaves the phase $\eta$ defined in (\ref{defeta1})
to be identified with $\eta (0)$. This identification is
extremely nontrivial mathematically, as it relates the
$\eta$-invariant in five dimensions to some rather
detailed properties of the eigenfunctions of the four dimensional
Dirac operator.
As mentioned previously, we have not rigorously proved
(although it is plausible - see the appendix)
that our phase $\eta$ converges to a well-defined limit as
the cutoff is taken to infinity. On the other hand,
$\eta (0)$ is defined in terms of an analytic continuation
which, while mathematically sound, may or may not
have the same physical content as our $\eta$
\footnote{In the work of Ball and Osborn \cite{eta}
similar results
are derived using Pauli-Villars rather than
$\zeta$-function regularization.}.
These issues clearly deserve further investigation.

\mysection{Reflections and Phases}

In this section we make an observation which simplifies the
form of the partition function (\ref{Z}).
We show that despite the possibility of complex phases in the
chiral determinant, in the absence of gauge anomalies
the partition function itself is real.
Our main observation is that the $\eta (0)$ part of the fermion
effective action changes sign when the gauge background
is reflected through any of the hyperplanes: $ \{ x_\mu ~=~ 0 \}$
-- in other words, it has odd parity.
This implies that all imaginary parts eventually cancel
in (\ref{Z}), leading to a real partition function.

Consider a background $\A$ and its reflection through
$x_4 = 0$:
\beq
A^*_\mu (x_i, x_4) ~\equiv~ A_\mu (x_i, -x_4).
\eeq
The five dimensional interpolations $^t \A$ and $^t \AS$
will also be reflections of each other.
We will specialize to the temporal gauge: $A_4 = 0$. We do not
lose any generality by doing so
if we are working in a model without gauge
anomalies, since in that case $\det$ is gauge invariant.
Note that the pure gauge actions
$S_{YM} [A]$ and $S_{YM} [A^*]$ are
identical.

We want to show two things:
\vskip .2in
\noindent (I) The phases of the determinant $\det$ in the backgrounds
$\A$ and $\AS$ are related by a minus sign. In other words,
\beq
\eta (0) \vert_{~^t A} ~=~ - ~ \eta(0) \vert_{~^t A^*}~~,
\eeq
which is equivalent to showing that there is a mapping
of $\lambda_n \rightarrow - \lambda_n$ when the gauge background
is reflected through $x_4 ~=~ 0$.

\vskip .2in
\noindent (II) The eigenvalues of $\DS_4$ are
invariant under $A \rightarrow A^*$.
This should be clear since the Euclidean Dirac operator can be regarded as
a  Hamiltonian, and the energy eigenvalues are invariant under
reflection of the gauge background. We will also see this explicitly
below by looking at eigenfunctions and their eigenvalues.
\vskip .2in

Given points (I) and (II), we can arrange the partition function in the
following way:
\beqa
\label{Z1}
Z ~&=&~ \sum_{  \{ A ~,~ A^*  \} } ~e^{ - S_{YM} [ A ]} ~ \det    \nonumber \\
     &=&~ \sum_{  \{ A  \}  }  ~ e^{ - S_{YM} [ A ]} ~ 2~ Re(  ~\det~  )
\nonumber \\
     &=&~ \sum_{  \{ A ~,~ A^*  \} } ~ e^{ - S_{YM} [ A ]}~Re(  ~\det~  ),
\eeqa
where first and third sums are over all configurations and
the second sum is only over the half of the possible configurations
which remain after modding out by the $Z_2$ reflection symmetry.
We have also used the fact that the sign of $\det$ is the same in the
$A$ and $A^*$ backgrounds. This follows from (II) and the tracking
of eigenvalues.

Now to the proof of (I).
The eigenvalue equation $\DS_5 \phi_n ~=~ \lambda_n \phi_n$
is as follows (we use the chiral basis for gamma matrices):
\beqa
\label{DE5}
\left[ +~ i \partial_5 u_n ~-~ i \partial_4 v_n ~+~
 D_i \sigma_i v_n \right] ~&=&~
 \lambda_n u_n \\
\left[ - ~i \partial_5 v_n ~-~ i \partial_4 u_n ~-~
 D_i \sigma_i u_n \right] ~&=&~
 \lambda_n v_n.
\eeqa
One can check that, corresponding to the original eigenfunction
of $\DS_5$ in the $^t A$ background,
\beq
\pmatrix{ u_n (x) \cr v_n (x) \cr },
\eeq
is an eigenfunction of $\DS_5$ in the
background $^t A^*$,
with
eigenvalue $ - \lambda_n$:
\beq
\pmatrix{  v_n(  x_i , -x_4, x_5) \cr
 u_n ( x_i , -x_4, x_5) \cr}.
\eeq
This shows that the eigenvalue spectrum of $\DS_5$ in the reflected
background is the negative of the original spectrum, which
is sufficient to prove (I).

We can see that this result also follows from the
functional integral form of the determinant
if we continue to work
in the temporal gauge. In this gauge a parity-like operation
relates the bases
$\{ \phi_n^L (x) \}$ and $\{ \phi_n^R (x) \}$ in backgrounds which
are related by a reflection through one of the hyperplanes
$\{ x_\mu ~=~ 0 \}$.

Given an eigenfunction (in the chiral basis
where $\gamma_5 ~=~ {\rm diag} \{ 1,-1 \}$)
of the Dirac operator in the background $A$:
\beq
\label{efn}
\phi_n (x) ~=~ \pmatrix{ \phi^R_n (x) \cr \phi^L_n (x) \cr },
\eeq
the following spinor function is an eigenfunction with the
same eigenvalue, but of the Dirac operator in the
background $A^*$:
\beq
\label{efn1}
\phi'_n (x) ~=~ \pmatrix{ -\phi^L_n (\vec{x}, -x_4) \cr
\phi^R_n (\vec{x}, -x_4) \cr}.
\eeq
(Note that this proves point (II) above.)
This similarity between the left and right chiral bases
in the two backgrounds
yields the following matrix relations:
\beqa
\label{crel}
C^L (A) ~&=&~ \{  C^R  ( A^* ) \}^* \nonumber \\
C^R (A) ~&=&~  \{ C^L ( A^* ) \}^*    ~.
\eeqa
Together, these imply that
\beq
det[~C^L (A)~] \cdot det[~C^R (A)~] ~=~
\{ det[~C^L (A^*)~] \cdot det[~C^R (A^*)~] \}^*,
\eeq
which is equivalent to (I).

We should note a limitation of the result:
the insertion of an operator into the sum in
(\ref{Z1}) (e.g. to compute an n-point correlator)
will in general destroy the reflection symmetry of the
terms in the sum. Therefore we can only compute
correlators using the weighting
$~Re(  ~\det~  )~$ if the operators
are themselves
invariant under at least one reflection, such as
\beq
\label{corr}
\langle ~ O(x_i, x_4) ~+~ O(x_i, - x_4) ~\rangle,
\eeq
where $O$ is a generic field operator.
This does not seem to be a significant limitation on
our ability to investigate issues of interest such as
chiral symmetry breaking, confinement or particle spectra.

\renewcommand{\thepage}{\arabic{page}}
\mysection{Conclusions}

We have proposed a formulation of chiral gauge
theory which should allow, in principle, the evaluation
of any quantity to a specified accuracy within
a finite computation. In order to avoid the problems
associated with lattice chiral fermions \cite{NN},
we employ
an interpolation of the original lattice gauge fields
to the continuum \cite{TH,FW,AG}.
All of the fermionic
aspects of the theory have been expressed here in terms
of the eigenvalues and eigenfunctions of the vector-like
Dirac operator, which can itself be implemented directly
on a lattice if desired.

The main focus of our investigation was
the chiral determinant defined
in terms of a regulated continuum functional integral.
When the gauge field
background is suitably well-behaved (e.g. resulting from lattice
interpolations like those described
in section 2), we find that an object with
all the desired properties can be extracted from the functional
integral as long as the mode cutoff $N$
used to regulate the integral
is kept sufficiently large.
For well-behaved background gauge fields,  any
violations of gauge invariance introduced by a hard mode
cutoff are confined to modes close to the cutoff.
These violations of gauge invariance do
not affect the complex phase information which characterizes
the chiral nature of the theory.
The order of limits that are necessary
for our regulator are as follows:
first, the continuum cutoff $N \rightarrow \infty$,
followed by the original gauge lattice spacing
$ a \rightarrow 0 $.

Our result for the chiral determinant is (from (\ref{detfin}))
\beq
\label{lastdet}
\det_{reg} ~=~  e^{i \eta} ~\prod_n ~\lambda_n~,
\eeq
where the phase $\eta$ is defined in (\ref{defeta1})
and the eigenvalues $\lambda_n$ are
described below (\ref{modes}).
Both $\eta$ (in the absence of anomalies
and in the limit $N \rightarrow \infty$)
and the eigenvalues are
gauge invariant functions of the background field.
In the appendix we give some evidence that suggests
that $\det$ as defined above will converge to a well-defined
value as $N \rightarrow \infty$, with its value mainly dependent
on modes with $n \lsim N_A$.
(\ref{lastdet}) is consistent with the
exact representation derived using $\zeta$-function regularization,
given the identification of the phases $\eta$ and $\eta (0)$
described in section 5.
One difference between the functional integral
formulation and the $\zeta$-function
regularization is that in the former the physical aspects
of the chiral determinant such as level crossing and the role of
high versus low frequency modes are more transparent.

Using our result for the chiral determinant we showed that
despite the sum over complex phases, the partition
function for a chiral gauge theory is real.
The simplified result for the partition function that we
derived in section 6 still requires the computation of the
non-anomalous phase $\eta$, as
\beq
Re( ~\det_{reg}~ ) ~=~ cos ( \eta )~ \prod_n \lambda_n ~.
\eeq
Therefore its main advantage is that
real rather than complex terms may be summed to
yield the final result.
It still remains a formidable technical problem
(although not one of principle)
to compute the phase $\eta$ on the lattice.

\vskip 1.0 in
\centerline{\bf Acknowledgements}
\vskip 0.1in

The author would like to thank R. Ball, S. Cordes, N. Evans,
P. Hernandez, G. Moore and S. Selipsky
for useful comments or discussions.
He especially thanks R. Sundrum for
emphasizing to him the possible utility of Fujikawa's
formalism for defining the chiral determinant, and to
G. Schierholz for pointing out an important error in a preliminary
version of this paper.
This work was supported under DOE contract DE-AC02-ERU3075.

\newpage
\mysection{Appendix}

In this appendix we discuss the behavior of Dirac
eigenfunctions for large $\lambda_n$  and implications
for the convergence properties
of the chiral determinant.

We begin by considering the two
component Dirac eigenvalue equation (\ref{qe1}), recopied
here for convenience.
\beqa
\label{diracA}
(q_4 ~+~ \lambda_n) u_n (q) ~&=&~  \int d^4k~ \left(
i A_4 (q-k) u_n (k) ~+~ A_i (q-k)
\sigma_i v_n (k) \right) \nonumber \\
(q_4 ~-~ \lambda_n) v_n (q) ~&=&~  \int d^4k~ \left(
i A_4 (q-k) v_n (k) ~+~ A_i (q-k) \sigma_i u_n (k) \right).
\eeqa
Consider the limit that $\lambda_n$
becomes arbitrarily large (in particular, we want
$n >> N_A$). We can see immediately that either
$v_n (q)$ or $u_n (q)$ must approach zero in this limit, depending
on the sign of $q_4$. Let us choose $q_4 < 0$ so that
$u_n (q)$ is nonzero.
The equations in (\ref{diracA}) then reduce to
\beq
(q_4 ~+~ \lambda_n) u_n (q) ~=~  \int d^4k~
i A_4 (q-k) u_n (k)
\eeq
up to corrections which vanish as $\lambda_n \rightarrow \infty$.
Note that we can set $A_4$ to zero by a
suitable gauge transform (which may
be different for different solutions due to the special
Lorentz frame we have chosen above).
Thus the eigenfunctions in this limit reduce to
free solutions, up to a gauge transform.
We therefore have
\beq
\label{CA}
\lim_ {m,n \rightarrow \infty}~~ C^L_{mn} ~=~ ( C^{R}_{mn} )^* ~,
\eeq
for arbitrary backgrounds $\A$ which are not necessarily
pure gauge.
Since the eigenvalues of (\ref{diracA}) are gauge invariant,
we also learn that
\beq
\label{limA}
\lim_{n \rightarrow \infty} ~\lambda^A_n ~=~ \lambda^0_n.
\eeq

There are several consequences of the results
(\ref{CA}) and (\ref{limA}).
Consider the Jacobian factor
\beq
det[ ~C^L~] \cdot det[~C^R~] ~=~ det[~ C^L (C^R)^T ~].
\eeq
(\ref{CA}) implies that for $N_A < ~ m, n~  < N - N_A$ the matrix
$ ~\{ C^L (C^R)^T \}_{mn}~$ is close to the identity $\delta_{mn}$,
and hence does not contribute to the phase $\eta$.
{}From (\ref{XA}) we know that
the diagonal corners of the matrix
$N - N_A < ~m, n~ < N$ are Hermitian.
Thus we expect that the non-anomalous phase
$\eta$ should roughly depend only on the $N_A \times N_A$
submatrices of $C^{L,R}$ and that
$\eta$ converges to a well-defined limit as $N \rightarrow \infty$
for fixed $a$.
This is intuitively plausible, since
it is mainly modes with $n \lsim N_A$ which are affected
by the presence of the
pre-regulated background field.

The result (\ref{limA}) is relevant to the existence
and convergence of  the limit
\beq
\label{ratio}
\lim_{N \rightarrow \infty} ~ \prod_n^{N} ~
{  \lambda^A_n  \over \lambda^{A'}_n}
\eeq
where $A$ and $A'$ are different background fields.
The existence of (\ref{ratio})
requires that very large eigenvalues are
essentially unperturbed by
a sufficiently well-behaved background $\A$ or $\A'$.
Without this property, the weighting factor of
gauge configurations in
(\ref{Z}) would be extremely difficult to compute
and could exhibit drastic
oscillations due to small changes in background field $\A$.
This question is actually also relevant to the
the fermion determinant in vector-like models like QCD,
and is not specific to chiral models.
Unquenched lattice QCD computations assume that
the ratio (\ref{ratio}) converges to its limit for
eiqenvalues of order the lattice spacing.

Closer examination of (\ref{diracA}) reveals that large eigenvalues
behave as
\beq
\lambda^A_n ~\sim~ \lambda^0_n ~+~ {\cal O} (A^2 / \lambda^0_n ).
\eeq
Using $\lambda^0_n \sim  ( n^{1/4} / L)$, where $L$ is the
size of our box, we have that
\beq
\label{sumA}
ln  \left(   \prod_n^{N} ~
{  \lambda^A_n  \over \lambda^{A'}_n} \right)
{}~\sim~ \sum^N_n ~    {\cal O}   \left(  {1 \over n^{1/2} } \right) .
\eeq
It is plausible that oscillations in the signs of
terms in the sum (\ref{sumA}) allow it to converge.
This would imply that (\ref{ratio}) is well-defined.


\end{document}